\documentclass[conference,10pt]{IEEEtran}

\usepackage{cite}
\usepackage{verbatim}
\usepackage{graphicx}
\usepackage[caption=false]{subfig}

\usepackage{enumitem}

\usepackage{fancyhdr}
\fancypagestyle{firstpage}{
  \fancyhf{}
  \fancyfoot[L]{{\bf [Cite As:]} R. Venkat, T. Divagar, T. Luo, and H.P. Tan, ``Participatory sensing for government-centric applications: a Singapore case study'', Technical Report, Institute for Infocomm Research, A*STAR, Singapore, 2014.}
}

\makeatletter
\def\ps@headings{%
	\def\@oddhead{\mbox{}\scriptsize\rightmark \hfil \thepage}%
	\def\@evenhead{\scriptsize\thepage \hfil \leftmark\mbox{}}
}
\begin{document}

\title{Participatory Sensing for Government-Centric Applications: A Singapore Case Study}

\author{
\IEEEauthorblockN{Ramgopal Venkateswaran, Thirumoorthy Divagar}
\IEEEauthorblockA{Raffles Institution, Singapore\\
Email: ramvenkat98@gmail.com, divagar@singnet.com.sg}\and
\IEEEauthorblockN{Tie Luo, Hwee Pink Tan}
\IEEEauthorblockA{Institute for Infocomm Research, A*STAR, Singapore\\
Email: \{luot,hptan\}@i2r.a-star.edu.sg}
}

\maketitle
\thispagestyle{firstpage}

\begin{abstract}
Singapore, an urbanized and populated country with high penetration of smartphones, provides an excellent base for citizen-centric participatory sensing applications. Mobile participatory sensing applications offer an efficient means of directing feedback to government agencies for timely identifying and solving problems of citizens' concern. While real deployments of such applications are on an uprising trend in Singapore, there is no concerted effort that studies the {\em user experience} of these applications. To fill this gap, we conduct a market study by analyzing the user reviews on the Google Play and Apple App Store for six major mobile crowdsourcing applications created by Singapore government agencies. This study was carried out for a period of 4 months during which we collected and analyzed 592 customer reviews. This was also supplemented by our personal use of the applications during the same period.

This paper presents the methodology and findings of this study, as well as our recommendations of what improvements that these applications could incorporate. We classify user reviews into 8 major concerns, and recommend 9 features to enhance the applications' utility. The recommendations are presented in terms of user interface, incentive, and publicity.
\end{abstract}


\section{Introduction}\label{sec:introduction}

Participatory sensing is a process that tasks mobile devices such as cellular phones to form interactive participatory sensor networks that enable public and professional users to gather, analyze and share local knowledge \cite{Gol0509}. 
There are several different usage models of participatory sensing, ranging from public contribution, in which individuals collect data in response to inquiries defined by a central entity or other individuals, to personal use and reflection, in which individuals log information about themselves \cite{estrin10inet}. This paper focuses on using participatory sensing as a civic tool to gather on-the-ground, citizen-initiated feedback that can be shared with relevant authorities as well as the rest of the public for efficient resolution of issues.

The high population density and mobile device penetration in Singapore presents great potential for participatory sensing to provision pervasive coverage without incurring cost of deploying instrumented sensors or specialized workforce. Therefore, it can possibly offer a cost-efficient and large-scale means of data collection and analysis across various industry sectors.

However, this potential has not been effectively exploited so far, according to a recent survey on real deployments of participatory sensing \cite{Til0313}. In the context of Singapore, while there are individual crowdsourcing applications designed for government agencies, the concept of participatory sensing has not gained much of the public's attention due to the fact that these applications pertain to specific, isolated issues such as monitoring litter, traffic or unsafe work practices. Convergence of these applications into a unified platform that accepts generic issues yet can categorize and direct the issues to the relevant government agencies, therefore, would foster a larger user base which is critical to the success of participatory sensing.


While much effort has been invested in developing participatory sensing applications, there is no concerted attempt to investigate user experience and lessons learnt from these applications. In this paper, we present a market research study we have conducted on extensive user reviews of a wide range of existing applications. We have also used all the applications in person for an extended period of time and incorporated our own experience into this study. Furthermore, based on the analysis of our findings, we recommend features and schemes that should enhance the application's utility and appeal to users.

\begin{figure*}[tb]
\centering
\includegraphics[width=0.8\linewidth]{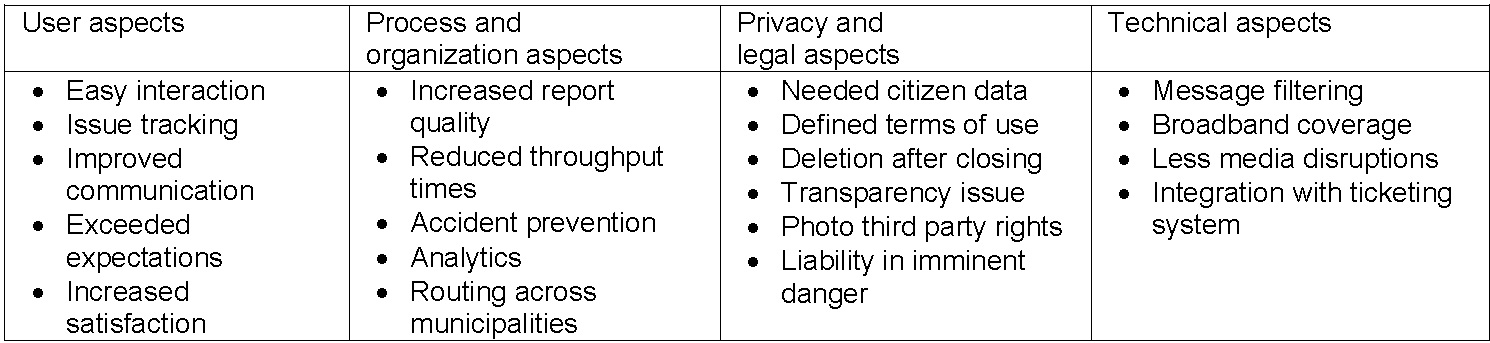}
\caption{Qualitative aspects of introducing mobile reporting.\cite{mobrep12}}
\label{fig:mobrep}
\end{figure*}

We also refer to some prior work conducted in various contexts as a guideline in planning and organizing our study as well as analyzing the market data. For example, \cite{mobrep12} derives some qualitative aspects of introducing mobile reporting, as shown in Fig.~\ref{fig:mobrep}.

The rest of the paper is organized as follows. We describe our methodology in Section \ref{sec:model}, and summarize our findings in Section \ref{sec:results}. We then present our recommendations in Section \ref{sec:rec}. Finally, we conclude with possible directions for future research in Section \ref{sec:conclusion}.

\section{Methodology}\label{sec:model}

This market research was conducted in three stages. The first stage involves background study and identifying prominent existing crowdsourcing applications to analyze. The second stage comprises collecting and analyzing user reviews of these applications at an extensive scale, as well as our personal usage of the applications. In the last stage, the user review data was categorized in a way that facilitates identifying larger and more general groups of concerns that consumers may have, by which we zoom out from micro to macro scope and provide recommendations that can be incorporated into both current and potentially future applications for various enhancements.

\subsection{Stage 1: Background Research and Identification of Relevant Applications}\label{sec:ups}

To ensure relevance to the development of participatory sensing applications, we had carefully selected a pool of mobile applications for our consumer review analysis. The basic criteria include (1) the applications were created by a Singaporean government agency rather than by private institutions or individual developers, because we theme around sustainable city development of Singapore and we also require the quality of the applications to be properly regulated, (2) the applications need to involve some form of crowdsourcing as a major element, and (3) the user base should be sizable and there are a considerable volume of customer reviews available, since these indicate the popularity of an application. Eventually, we were able to include in out study not only the most favored applications in terms of user reviews but also the least liked applications. This allows us to understand both the desirable features of consumers as well as their turnoffs. On our final list are six crowdsourcing applications: MyEnv\cite{myENV}, CleanLah\cite{CleanLah}, and WeatherLah\cite{WeatherLah} by the National Environmental Agency (NEA), Snap@MOM\cite{SnapatMOM} by the Ministry of Manpower (MOM), MyTransport\cite{MyTransport} by the Land Transport Authority (LTA), and Police@SG\cite{PoliceSG} by the Singapore Police Office.

\subsection{Stage 2: Verification, Analysis and Categorization of Customer Reviews}\label{sec:algo}

The purpose of this stage was to gain insights into the public opinion on the pros and cons of current crowdsourcing applications. Besides analyzing existing user reviews, it is also a key to gain first-hand experience. Therefore, we downloaded them into our own phones, Samsung Galaxy Grand, Motorola MB525, and iPhone 4, which run the two most popular mobile OS (Android and iOS), and used them in person for 4 consecutive months. 

We created a database for each of the participatory sensing applications under investigation. Then, we read and verified all the user comments on the Playstore and Appstore for these applications, and populated the database by importing the comments into the database separately as per whether they are positive or negative comments.

To ensure the relevancy and accuracy of user comments, we also verified them carefully based on the latest updated version of the pertaining application. For example, if a user commented that there was no way to turn off notifications but there actually was, this comment will instead be categorized as confusing interface design because the user did not manage to locate the notifications toggle button. If a comment is based on an early version of the application that does not have the notification toggle button, while the latest version does have, the comment will be deemed as invalid or outdated and not imported into the database. The above verification process also took into account the different contexts of the comments in a case-by-case manner. The comments were finally categorized into broader groups of user concerns for each particular application, with a number indicating the size of each group. 

In total, we have collected and analyzed 592 user comments on Google Play and Apple App Store over a period of four months (August--December 2013). Due to the casualness of users writing reviews, we filtered out many uninformative or invalid comments, and eventually selected 281 useful reviews to compile our findings.\footnote{In some not-so-common cases, a user commented on different aspects of an application, in which case each aspect was counted as a separate comment and categorized accordingly.} Furthermore, we have also augmented this user review database with our own personal use of the applications on both Android and iOS devices over the same period of four months.

\subsection{Stage 3: Cross-Application Generalization}

In this stage, we generalize the above application-specific categories to wider user concerns across all these applications. The purpose was to create a guideline for crowdsourcing application designers and developers to refer to, particularly for better user experience and a broader outreach. Ultimately, we provide an extensive list of recommendations that are applicable to a wide range of crowdsourcing applications, covering several important aspects in system design, implementation, and deployment.

\begin{table*}[tb]
  \caption{Summary of User Reviews by Category}\label{tab:reviews}\vspace{-1mm}
\begin{center}
  \begin{tabular}{ l | c | c | c } \hline
{\bf Category} & \multicolumn{2}{|c|}{\bf No. of User Reviews} & \% of Total \\
 & Positive & Negative & \\ \hline
Accuracy and Usefulness of Information   &  32 & 52 & 29.9 \\
User Interface Design & 20 & 48 & 24.2 \\
Compatibility Issues, Responsiveness and Robustness & 3 & 56 & 21.0 \\
Quality of Alerts, Notifications and Widgets & 4 & 14 & 6.41 \\
Government’s Response and Further action & 4 & 10 & 4.98 \\
Purpose of Application & 1 & 13 & 4.98 \\
Quality of Application Updates & 8 & 0 & 2.85 \\
Network Connection Issues & 0 & 5 & 1.78 \\
Miscellaneous & 2 & 9 & 3.91 \\ \hline
{\bf Total} & 74 & 207 & 100 \\ \hline
  \end{tabular}
\end{center}
\end{table*}

\section{Findings}\label{sec:results}

This section summarizes our findings. We categorize the 281 user reviews into 8 major concerned areas as well as a miscellaneous category, as shown in Table.~\ref{tab:reviews}.

\subsection{Accuracy and Usefulness of Information}
This is an often mentioned user concern found in 29.9\% of all the user reviews. A lack of reliable and accurate information is a substantial and non-trivial issue that faces crowdsourcing applications. There needs to be a good mechanism that can determine the validity and trustworthiness\cite{sew14secon} of crowdsourced information and sieve out fake and malicious information. There is also a need for sufficient deterrents to preemptively prevent users from posting false data. Besides, the usefulness of information often differs from individual to individual; a mechanism that allows a user to personalize the application such that he only see information useful to himself would therefore be a great advantage.

\subsection{User Interface Design}
User interface design is the second largest category due to its permeating presence throughout every application. This was corroborated by our findings, where 24.2\% of reviews pertained to user interface design. There are three main subcategories (see Table.~\ref{tab:UI}) that users are specifically concerned about: ease of use, aesthetic appeal, and personalization.

\begin{table}[ht]
  \caption{User Reviews about User Interface Design}\label{tab:UI}\vspace{-2mm}
\begin{center}
  \begin{tabular}{ c | c | c | c } \hline
{\bf Aspects of UI} & Ease of Use & Aesthetic Appeal & Personalization \\ \hline
{\bf No. of Reviews} & 32 & 33 & 3 \\ \hline
  \end{tabular}
\end{center}
\end{table}

\begin{figure*}[t]
\centering
\subfloat[WeatherLah's map presentation allows users to view weather throughout the whole island of Singapore and has been complimented by many user reviews.]
{\label{fig:WeatherLah}\includegraphics[width=0.3\linewidth]{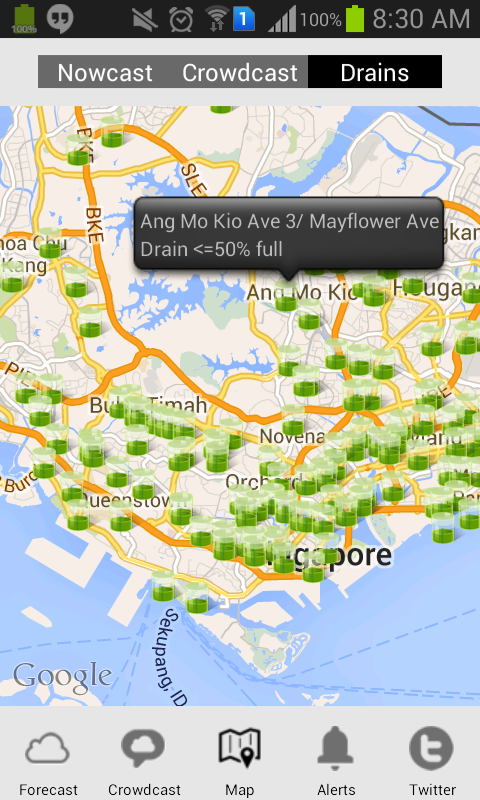}}\hfil
\subfloat[Snap@MOM has vibrantly colored illustrations in a simple user interface.]{\label{fig:snap}
\includegraphics[width=0.3\linewidth]{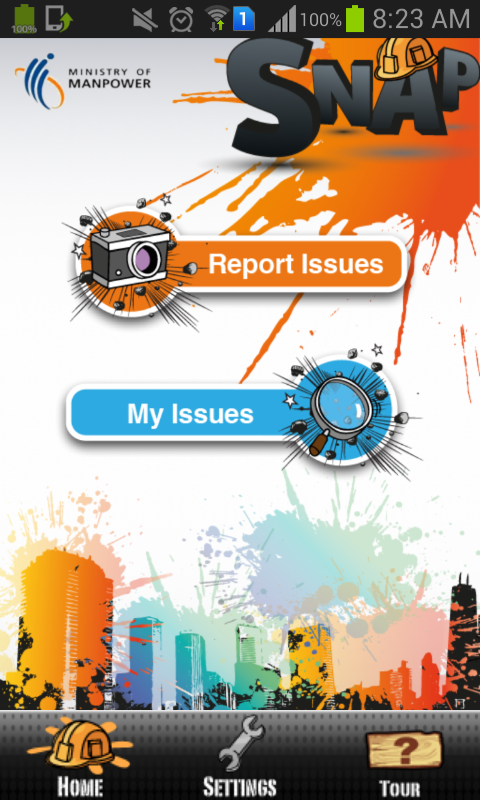}}\hfil
\subfloat[MyEnv has the option to allow users to see different screens upon opening the application, according to their preference.]{\label{fig:myenv}
\includegraphics[width=0.3\linewidth]{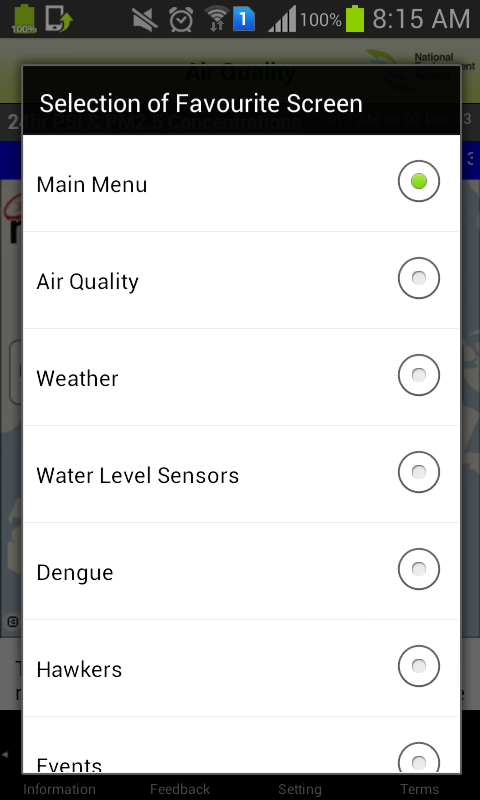}}
\caption{Representative user interfaces.}\label{fig:personUI}
\end{figure*}

\begin{itemize}[leftmargin=1.5em]
\item {\it Ease of Use}:
It is important to users that ``the client application of a mobile reporting service should possess high usability and enable easy interaction to outmatch traditional channels of communication'' \cite{mobrep12}. It is often helpful for applications to provide users with enough options to cater for different individuals' various needs; however, at times, a user has to work through a mesh of different options to locate what he really needs. As a result, words such as ``confusing'' and ``disorganized'' appear frequently in the reviews of certain applications. Reviews also suggest that users tend to favor macro-scale pictorial representation that readily gives an intuitive big picture as well as supplements detailed specific information; Fig. \ref{fig:WeatherLah} gives such an example, where WeatherLah's map presentation gained compliments from many user reviews.
\item {\it Aesthetic Appeal}:
While solid functionality is definitely a necessity, aesthetic appeal draws the user to the application---about one third of user comments about UI concern with aesthetic appeal, highlighting its importance. Users prefer applications that use bold and vibrant colors rather than plain ones; yet these applications should not be cluttered either. A positive example in this respect is Snap@MOM (see Fig. \ref{fig:snap}). In addition, the UI design also needs to take into account the tastes of the target user group of the application. MyEnv\cite{myENV} is such an example where Android users were dissatisfied with an interface ported from iOS while they preferred a Holo-style design.
\item {\it Personalization}:
Applications should be customizable to users' unique preferences---while there are a great number of applications which allow users to select parameters for the notifications and alerts they receive, the main interface, however, is sometimes rather rigid in that users are not able to personalize it according to their preferences. Incorporating this feature would make the application simpler to use. MyEnv represents one of the applications that allows users to select the main UI they want to use; see Fig. \ref{fig:myenv}.
\end{itemize}

\subsection{Compatibility Issues, Responsiveness and Robustness}
From our own experience on the Motorola MB525 Android phone which is a relatively old model, many of the applications, particularly MyEnv and WeatherLah, often ran very slowly or even suddenly crash. These compatibility, responsiveness\cite{ictc12}, and robustness issue of such applications was concerned by 5\% of user reviews and we deem it should not be taken lightly and is worth designers and developers' effort to improve.

\subsection{Quality of Alerts, Notifications and Widgets}
While alerts, notifications and widgets enable a user to receive important information on the go, 6.41\% of all user reviews and 6.76\% of all negative reviews mentioned that many notifications are not sufficiently relevant to their needs and some are way too frequent and annoying. Therefore, in order to accommodate different user needs, notifications should be highly customizable to cater to users' priorities and convenience. Some users also encountered difficulty in navigating through a complex interface to toggle notifications, hence hinting toward the need for an intuitive user interface design.

\subsection{Government's response and further action}
The applications CleanLah and Snap@MOM, which function by relying on participants feedback, received a large percentage (35.7\% and 69.2\% for these two applications, respectively) of user comments that mentioned government's response as a critical factor. These reviews typically concerned with the speed and effectiveness of response as well as whether users are informed of the follow-up to their feedback. Applications thus need to be able to channel information in both directions, from users to government as well as from government to users. Information updates about the resolution of issues, as well as statistics that illustrate adequate responses, would therefore be desirable.

\subsection{Other User Concerns}

Besides the above 5 user concerns which constitute 86.5\% of user reviews, the following 3 categories add up to only 9.6\% of user comments. However, we deem them to be worth considering due to their relevancy to user experience and based on our own personal usage as well.

\begin{itemize}[leftmargin=1.5em]
\item {\it Purpose of Application}: Some users were unclear of the purpose of certain applications (MyEnv, MyTransport, Police@SG). This is likely because they were bombarded with a lot of information not well organized but cluttered in the application.
\item {\it Quality of Application Updates}: User reviews indicated that the application should not stagnate but improve over time by releasing timely updates. On the other hand, the updates should not compromise the smooth functioning of the application and hence should be released only when changes are nontrivial.
\item {\it Network Connection Issues}: Some users experienced problems of Internet connections when accessing or using the applications, especially when they are traveling in vehicles. This suggests that developers may need to consider other communication options such as SMS or MMS.
\end{itemize}

\section{Recommendations}\label{sec:rec}

In this section, we provide relevant recommendations which would improve participatory sensing applications over the issues expressed in the user reviews. When preparing these recommendations, we took into account public user reviews, our personal usage of the applications, and the qualitative analysis of participatory sensing from Fig.~\ref{fig:mobrep}. These recommendations are classified into 3 broad categories, namely user interface design, incentive, and publicity.

\subsection{User Interface}

Existing UI can be improved in terms of clarity, accuracy, and accessibility.

\begin{enumerate}[leftmargin=1.5em]
\item{\textbf{Dual-Interface System}} \par Participatory sensing applications that source for user feedback rarely allow for two-way communication in that government agencies are not able to communicate with users, which is however important because involving government agencies to communicate to users within the application itself would facilitate quick dissemination of information as well as attracting a larger user base and user confidence in the application.
A system that allows government agency to log in (probably with a separate interface, such as Snap@MOM illustrated by Fig. \ref{fig:snap-login}) would make this viable. This would also bolster follow-up actions such as government responses, and allows government agencies to play a role in, e.g., severity rating.
\begin{figure}[ht]
\centering
\includegraphics[width=0.6\linewidth]{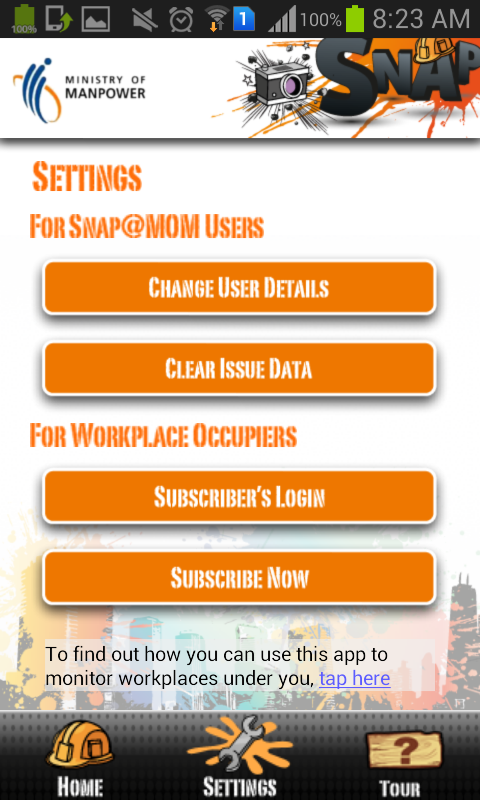}
\caption{Snap@MOM allows users of different roles to log in with different permissions.}
\label{fig:snap-login}
\end{figure}
\item{\textbf{Quality and Validation}} \par As the quality of contribution\cite{qcs13dcoss} is critical to participatory sensing, validation becomes an important way to prevent users from being misled by false or inaccurate reports from other users. A possible solution is to design truth-finding or information-validation algorithms using machine learning techniques. Another way is to allow users to rate each other's reports in terms of quality or validity.

\item{\textbf{Alternative Communication Channels}} \par Most participatory sensing applications require an Internet connection to send user feedback to the application database. However, a stable Internet connection is not always available when users are on the go. Therefore, having an option in the application to send the information through messaging via telco networks would be a solution worth consideration. To this end, a special telco number needs to be registered with a telco.
\item{\textbf{Widgets}} \par This recommendation specifically pertains to Android phones. Widgets allow for user-specific information to be easily accessed and quickly seen by the user on the home screen of his phone. Users can prioritize on which type of information to be displayed on the widget and specify what frequency the information should be updated. Several applications have an option to activate widgets, but lack the flexibility of personalizing them to users' own needs and interests, e.g., by relating to a certain residential area or a certain type of issues.
\item{\textbf{Severity Rating}} \par The owner of a post would be allowed to assign a severity value so as to indicate the importance or urgency of the pertaining problem. On the other hand, as it only reflects the subjective view of the owner, the application may also allow other users to add their ratings in order to calculate an average which bears more objectiveness. In addition, the government agency may also be allowed to rate the issue for greater conformity and less subjectivity. An aggregation algorithm could be designed to take into account different weights of different users and government agencies.
\item{\textbf{Pictorial Representation}} \par A macro-scale pictorial representation enhances clarity by providing users who do not want to read text-based details with a quick and easy way to get an overall sense, such as all problem occurrences in the region together with their nature and severity. Fig.~\ref{Fig:pictorial} illustrates this idea. Also, problems may be color coded (e.g. according to severity rating) or with different icons (e.g., garbage representing a cleanliness problem while a vehicle representing a traffic problem), and users may be able to set a filter to see problems of his interest only.

\begin{figure}[ht]
\begin{center}
\includegraphics[width=\linewidth]{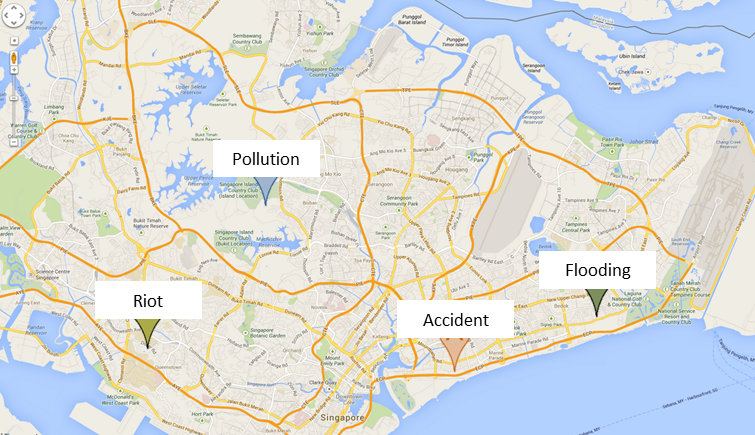}
\caption [Illustration of color-sensitive graph coloring] {
\label{Fig:pictorial}{Pictorial representation as in a map.}}
\end{center}
\end{figure}
\end{enumerate}

\subsection{Incentive}

Incentives are key to make participatory sensing applications successful because these applications completely rely on the public users' participation in order to collect data or feedback.

\begin{enumerate}[resume,leftmargin=1.5em]
\item{\textbf{Incentive mechanisms}} \par Incentive mechanisms are an effective and theoretically proven tool to motivate users to make contributions in participatory sensing. There is a rich literature dwelling on this; for example \cite{infocom14prof} proposed an all-pay auction based incentive mechanism and could be referred to.

\item{\textbf{User ranking}} \par Employing a weekly/monthly/yearly leaderboard system would provide recognition to committed users and serve as motivation to continually posting. Each post should earn points depending on its spatiotemporal property, importance, urgency, etc. The leaderboard thus may comprise those who have earned the most number of points within a certain period of time. In a sense, it also adds some gamification element and fun competition to the application to appeal to users.
\end{enumerate}

\subsection{Publicity}

Publicity can be enhanced by capitalizing prominent existing platforms.
 
\begin{enumerate}[resume,leftmargin=1.5em]
\item{\textbf{Social Media}} \par The advent of Facebook has made waves around the world with 1.2 billion users in just 8+ years. This predisposed popularity of social media over the world populace presents social media as an effective platform to boost a larger user base for crowdsourcing. Using social media like Facebook and Twitter as a publicity mechanism would provide users of participatory sensing applications an option to share their posts on particular issues on other popular channels. Furthermore, enabling the penetration of information in this way also helps gain user support from a larger community.
\item{\textbf{Accompanied Website}} \par A website specifically designed for a mobile application can serve to publicize the application as well as incentivize users to use the application by providing statistics, figures and pictures, without cluttering up the phone interface. It allows users to better understand the application's purpose and features. The site also helps involve users to a deeper extent, by including a page where users can share feedback about the application, and the feedback can be channeled to the developers. In fact, it becomes a vehicle for developers to crowdsource ideas and suggestions from users to incorporate into the applications.
\end{enumerate}

\section{Conclusion}\label{sec:conclusion}

In this paper, we present an extensive market study we have conducted on government-centric participatory sensing applications in Singapore, with the purpose of understanding the desirable features for sustained use of these applications. Our study has identified 8 major user concerned areas and provided 9 relevant recommendations for both existing and future participatory sensing applications.

Further scope for research would include the use of machine learning techniques to facilitate analyzing the data sets, which was carried out manually in this work. That way, we should achieve much greater efficiency and be able to analyze a much larger set of customer reviews as well as other data sources. 



\bibliographystyle{IEEEtran}
\bibliography{issnip2014}
\end{document}